\documentclass[twocolumn,secnumarabic,amssymb, amsmath, nofootinbib,tightenlines,
nobibnotes, aps, prl]{revtex4}

\usepackage{graphicx}
\usepackage{dcolumn}
\usepackage{bm}

\newcommand {\be}{\begin{equation}} 
\newcommand{\ee}{\end{equation}}    
\def\ddt{\frac{\partial}{\partial t}}
\def\dds{\frac{\partial}{\partial s}}
\def\dds1{\frac{\partial}{\partial s_1}}

\def\d{d\kern-0.8 ex\vrule height 1.3 ex depth-1.24 ex width 0.7 ex
\kern 0.15 ex}
\def\D{D\kern-1.7 ex\vrule height .87 ex depth-0.8 ex width 0.7 ex
\kern 0.95 ex}

\begin{document}


\title{Comment on 'Alfv\'en Instability in a Compressible Flow' [Phys. Rev. Lett. {\bf 101}, 245001 (2008)]}

\maketitle

The Alfv\'en wave has been a popular subject in various studies in
the past and, in particular, in studies dealing with the heating of
the solar corona and the consequent generation or acceleration of
the solar wind. However, to find a widespread source for the wave,
appears to be a major problem. In a recent paper~\cite{tar}, it is
claimed that a new MHD instability of the Alfv\'en wave has been
found: incompressible Alfv\'en modes propagating in a compressible
spatially varying flow were apparently exponentially amplified.
Bearing in mind the importance of the Alfv\'en wave and the possible
consequences in case of such a widespread source for its generation,
we here reexamine the model used in Ref.~\cite{tar}.

We have found that the results obtained in the work \cite{tar} are
wrong. This  because  the `equilibrium' used in Ref.~\cite{tar} is
only assumed, instead of being  self-consistently determined from
the equations, and it does not exist. Below, we explain this in more
detail.

The general starting equations should read
\begin{eqnarray}
\rho \left(\ddt + \vec V \cdot\nabla \right) \vec
V\!\!&\!\!=\!\!&\!\! -\nabla \left(p
+\frac{B^2}{2 \mu_0}\right) + \frac{(\vec B\cdot\nabla)}{\mu_0} \vec B, \label{e1} \\
\frac{\partial \vec B}{\partial t}&=& \nabla\times (\vec V\times \vec
B), \label{e2} \\
\frac{\partial \rho}{\partial t} +  \nabla\cdot ( \rho \vec V)&=&0.
\label{e2a}
\end{eqnarray}
In Ref.~\cite{tar} it is assumed that $\vec B_0=B_0\vec e_z$, while $B_0=const.$, and
the equilibrium velocity is $\vec u_0=u_0(z)\vec e_z$, while the
pressure term is omitted. The model thus implies a spatially
accelerated/decelerated equilibrium (or background) plasma, although
the source for this effect is missing.

In  Ref.~\cite{tar} only the continuity equation (\ref{e2a}) is used
to describe the equilibrium,  yielding
\be
\rho_0(z) u_0(z)=c_1, \label{con} \ee
where $c_1$ is a constant.  Taroyan  makes a mistake here by {\em
assuming} an equilibrium velocity, i.e.,  by setting it 'to have a
step function profile', instead of obtaining it from Eq.~(\ref{con})
and from the  momentum equation  Eq.~(\ref{e1}), when it is used to
describe the equilibrium, or some other form of the momentum. We
stress that the spatial variation of the plasma velocity is well
known in the models dealing with the solar wind (as clear from the
references cited in \cite{tar}). However, in those models it follows
{\em self-consistently} from the momentum and continuity equations.

As a matter of fact, it is seen that without the pressure term, for
a time-independent equilibrium/background plasma, the right-hand
side in Eq.~(\ref{e1}) vanishes so that
\be
\frac{\rho_0}{2} \frac{\partial u_0^2}{\partial z}\equiv c_1
\frac{\partial u_0}{\partial z}= 0. \label{e3} \ee
However, this contradicts the condition (\ref{con}) where $u_0$ is assumed
to be depending on $z$. Clearly, one way out of this  is  to keep  the pressure term in Eq.~(\ref{e1})
when it is used to describe the equilibrium. This in fact is
equivalent to taking into account the compressibility in the
equilibrium, as done inconsistently in Ref.~\cite{tar}. This yields
the second condition for the equilibrium:
\be
u_0(z)^2+ 2 c_s^2 \ln \rho_0(z)=c_2=const. \label{sc} \ee
Here, $c_s^2=\kappa T/m$. Eq.~(\ref{sc}) is to be used together with Eq.~(\ref{con}) in order to
self-consistently determine the possible profiles for $\rho_0(z)$
and $ u_0(z)$.

In other words, the equilibrium plasma flow considered by Taroyan contains a step
function velocity, i.e.\ a (steady) shock. This is all right as long as the
 Rankine-Hugoniot conditions (all of them!) are satisfied across the discontinuity
 (see e.g.\ \citet{GP}, Chap.~4, p.~170). The Rankine-Hugoniot condition
 following from the continuity equation is the one considered by Taroyan,
  viz.\ $\lbrack\!\lbrack\rho_0u_0\rbrack\!\rbrack=0$, i.e., $\rho_0u_0$
  has to be constant across the discontinuity. However, the momentum equation
   also yields a Rankine-Hugoniot jump condition. In the simplified set-up
   considered, this condition reduces to $\lbrack\!\lbrack\rho_0u_0^2\rbrack\!\rbrack=0$,
    which is equivalent to the condition (\ref{e3}) mentioned above. Hence,  the combination of the two condition results in
     $\lbrack\!\lbrack u_0\rbrack\!\rbrack=0$, i.e.,\ if the pressure is
      ignored, the equilibrium velocity can not jump. But if the pressure  is
      kept then both Eqs. (\ref{con}),~(\ref{sc}) must be used.

Hence, the equilibrium velocity  cannot be just
'assumed' to have an arbitrary step profile. The step has to satisfy all the
Rankine-Hugoniot conditions. Moreover, to obtain a `steady' shock, the shock
speed should be zero, which is clearly not the case in the plasma flow
considered by Taroyan.  The results obtained in Ref.~\cite{tar} are thus
wrong and the conclusions should be disregarded.


Acknowledgement: the  results   are  obtained in the framework of
the projects G.0304.07 (FWO-Vlaanderen), C~90205 (Prodex~9), and
GOA/2004/01 (K.U.Leuven).

\vspace{.5cm}

\noindent{\bf  J. Vranjes and S. Poedts}\\
\noindent K. U. Leuven, Center for Plasma Astrophysics,
Celestijnenlaan 200B, 3001 Leuven,
 Belgium, and Leuven Mathematical Modeling and Computational Science Center
 (LMCC)

\vfill

\end{document}